\documentclass{appolb}

\usepackage[T1]{fontenc} % The T1 font encoding is an 8-bit encoding and uses fonts that have 256 glyphs, fonts EC

\usepackage{graphicx}    % graphicx package included for placing figures in the text
\usepackage{amsmath}
\usepackage{rotating}
\usepackage{bm}          % bold math
\usepackage{amssymb}
\usepackage{amsmath}

\begin{document}
\title{PROPERTIES OF SUPERHEAVY ISOTOPES $Z=120$ AND ISOTONES $N=184$ WITHIN THE SKYRME-HFB
MODEL\thanks{Presented at the XXIV Nuclear Physics Workshop "Marie and Pierre Curie",\\
 Kazimierz Dolny, Poland, September 20--24, 2017.}%
}
\author{A. Kosior, A. Staszczak
\address{Institute of Physics, Maria Curie-Sk{\l}odowska University, Lublin, Poland}
\\
\bigskip
Cheuk-Yin Wong
\address{Physics Division, Oak Ridge National Laboratory, Oak Ridge, TN USA}
}
\maketitle

\begin{abstract}

We study the nuclear properties of even-even superheavy $Z$=120  isotopes and $N$=184
isotones with the Skyrme Hartree-Fock-Bogoliubov (HFB) approach. Within this model we
examine the deformation energy surfaces and two paths to fission: a reflection-symmetric
path with elongated fission fragments (sEF) and a reflection-asymmetric path corresponding
to elongated fission fragments (aEF).
Furthermore, we explore the energy surfaces in the region of very large oblate deformations
with toroidal nuclear density distributions.
While the energy surfaces of toroidal  $Z$=120  isotopes and $N$=184 isotones do not
possess energy minima without angular momenta, local energy minima (toroidal high spin
isomeric states) appear for many of these superheavy nuclei with specific angular momenta
about the symmetry axis.
We have theoretically located the toroidal high spin isomers (THSIs) of $^{302}$Og$_{184}$, 
$^{302}120_{182}$, $^{306}120_{186}$, and $^{306}122_{184}$.
%The occurrence of THSIs appears to be rather common in the superheavy nuclei region.
\end{abstract}

\section{Introduction}

Nuclei with excessive charges have a tendency to distribute the density in the oblate
configuration in the shape of a biconcave disks or a toroid \cite{Sta09,Kos17,Sta17}.
The additional presence of a large angular momentum about the symmetry axis enhances
the stability of a toroidal density distribution.
Previously, we examined the energy surfaces of even-even superheavy $Z=120$ isotopes and $N=184$
isotones in extremely prolate and oblate shapes, without and with an angular momentum \cite{Kos17,Sta17}.
We found that even though toroidal density distributions frequently appear in constraint HFB
calculations, there is no local toroidal energy minimum when there is no angular momentum
in this region of $Z$ and $A$.  However, under the constraint of an angular momentum about
the symmetry axis,  toroidal high spin isomers (THSIs) show up as local energy minima in
$^{304}$120$_{184}$ when it is endowed with specific angular momenta \cite{Sta17}.
In the present contribution, we continue our examination of the energy surfaces of other
even-even superheavy $Z=120$ isotopes and $N=184$ isotones and search for THSIs under
the constraint of angular momenta about the symmetry axis. We find many local energy
minima, THSI states, in the even-even neighborhood of $^{304}$120, indicating the common
presence of toroidal high-spin isomer in the superheavy region.

\section{Skyrme-HFB Model}

In our method, we use the constrained and/or cranked Skyrme Hartree-Fock-Bogoliubov (HFB)
approach, which is equivalent to minimization of the Skyrme energy density functional
$E^{tot}[\boldsymbol{\bar{\rho}}]$, with respect to the densities and currents under
a set of constraints (see Ref.~\cite{Sta17} and references cited therein).
Taking the method of Lagrange multipliers, we solve an equality-constrained
problem (ECP):
\begin{equation}
\left\{
\begin{array}{l}
\displaystyle\min_{\boldsymbol{\bar{\rho}}} E^{tot}[\boldsymbol{\bar{\rho}}]\\
         \mbox{subject to: } \displaystyle\langle \hat{N}_{q} \rangle= N_{q},\quad (q=p,n),\\
\phantom{\mbox{subject to: }}\displaystyle\langle \hat{Q}_{\lambda\mu} \rangle= Q_{\lambda\mu},\\
\phantom{\mbox{subject to: }}\displaystyle\langle \hat{J}_{i} \rangle= I_{i}, \quad (i=x,y,z),
\end{array}
\right. %\nonumber
\label{eq:1}
\end{equation}
where the constraints are defined by the average values $N_{p,n}$ of the proton
and neutron particle-number operators $\hat{N}_{p,n}$, the constrained values
$Q_{\lambda\mu}$ of the mass multiple moment operators $\hat{Q}_{\lambda\mu}$,
and the constraint components of the angular momentum vector $I_{i}$.
The above ECP equations are solved using an augmented Lagrangian method \cite{Sta10}
with the symmetry-unrestricted code HFODD \cite{hfodd}, which uses the basis expansion
method utilizing a three-dimensional Cartesian deformed harmonic oscillator (h.o.) basis.
The basis was composed of the 1140 lowest states taken from the $N_{0}=26$ h.o. shells.
In the particle-hole channel, the Skyrme SkM* force \cite{Bar82} was applied
and a density-dependent \textit{mixed} pairing \cite{Sta17} interaction in
the particle-particle channel was used.

We use the constrained Skyrme-HFB approach when we try to establish the region of
the quadrupole deformation with the spherical and toroidal density distributions.
The cranked Skyrme-HF model (without pairing correlations) was applied to locate
the THSIs under the constraint of an angular momentum about the symmetry axis.

%\section{Results}
\section{Energy Surfaces of $Z$=120 Isotopes and $N$=184 Isotones}

\begin{figure}[htb]
\begin{center}
\includegraphics[width=0.75\textwidth]{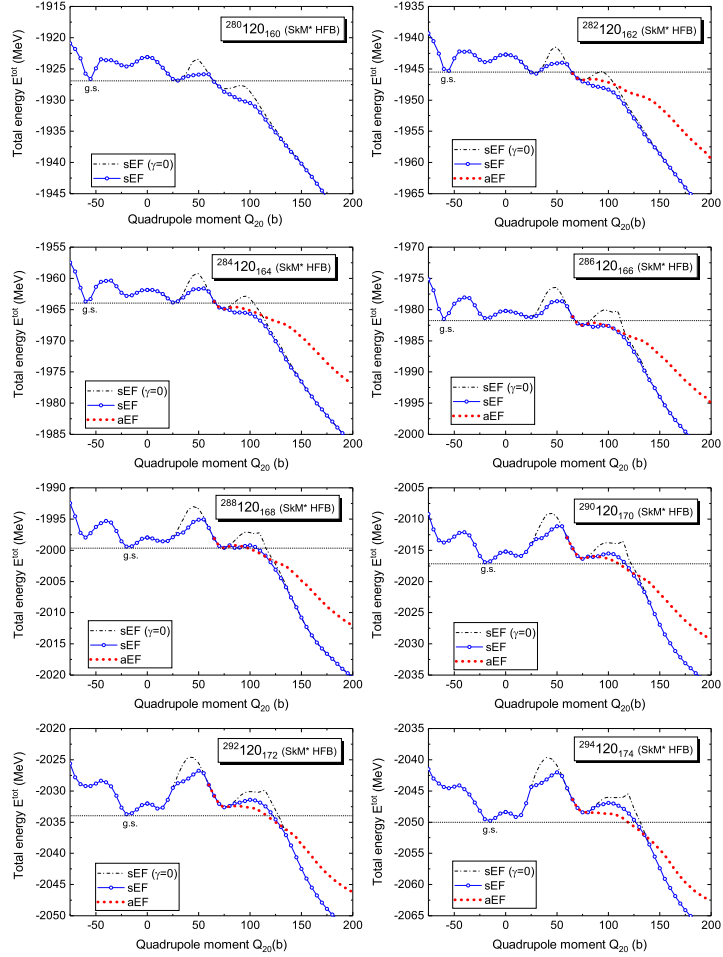}
\caption{\label{fig:1} (Color online) Total HFB energy as a function of the quadrupole
moment for the even-even $Z=120$ isotopes with number of neutrons from $N=160$ to $174$.
The open circular points (blue color) and short dashed (red) lines show the symmetric (sEF)
and asymmetric (aEF) elongated fission pathways, respectively. The axially symmetric
sEF ($\gamma =0^{\circ}$) fission pathways are marked by the dash-dot curves.
%The horizontal short dot lines represent ground state (g.s.) energies of these nuclei.
}
\end{center}
%\vspace*{-0.7cm}
\end{figure}

\begin{figure}[htb]
\begin{center}
\includegraphics[width=0.75\textwidth]{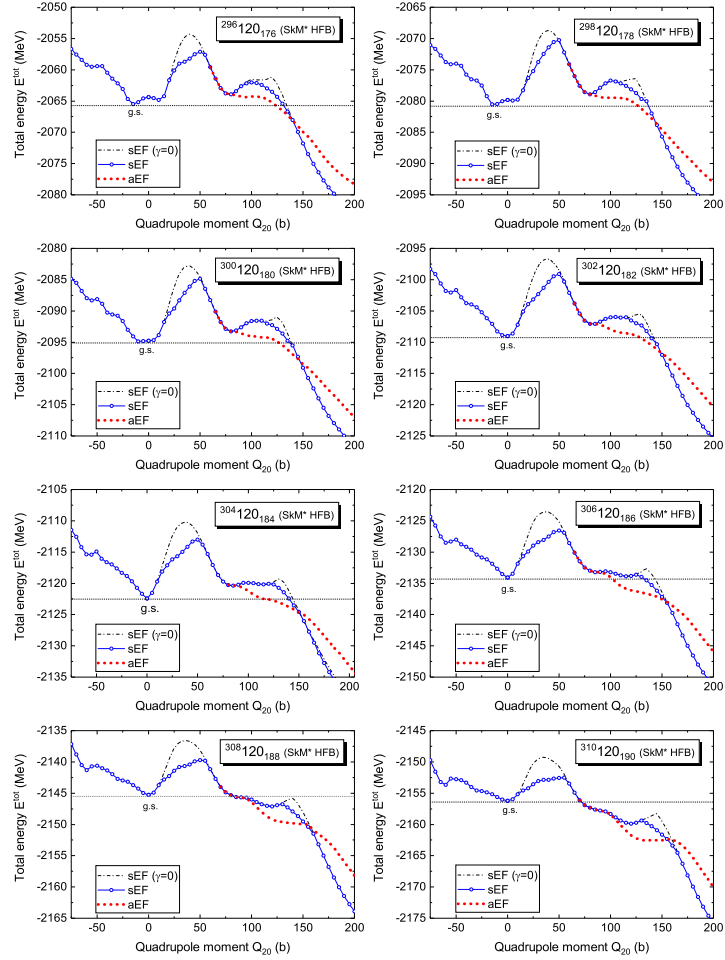}
\caption{\label{fig:2} (Color online) The same as Fig.~\ref{fig:1}, but for number of
neutrons from $N=176$ to $190$.
}
\end{center}
%\vspace*{-0.9cm}
\end{figure}

Figs.~\ref{fig:1} and ~\ref{fig:2} present the total HFB energy of even-even superheavy
$Z=120$ isotopes with number of neutrons from $N=160$ to $190$. For each isotope except
$^{300}$120$_{160}$, one can see two paths leading to fission: a reflection-symmetric
path with the elongated fission fragments (sEF) (open circles) and a reflection-asymmetric
path with the elongated fission fragments (aEF) (dashed curves).

\begin{figure}[htb]
\begin{center}
\includegraphics[width=0.65\textwidth]{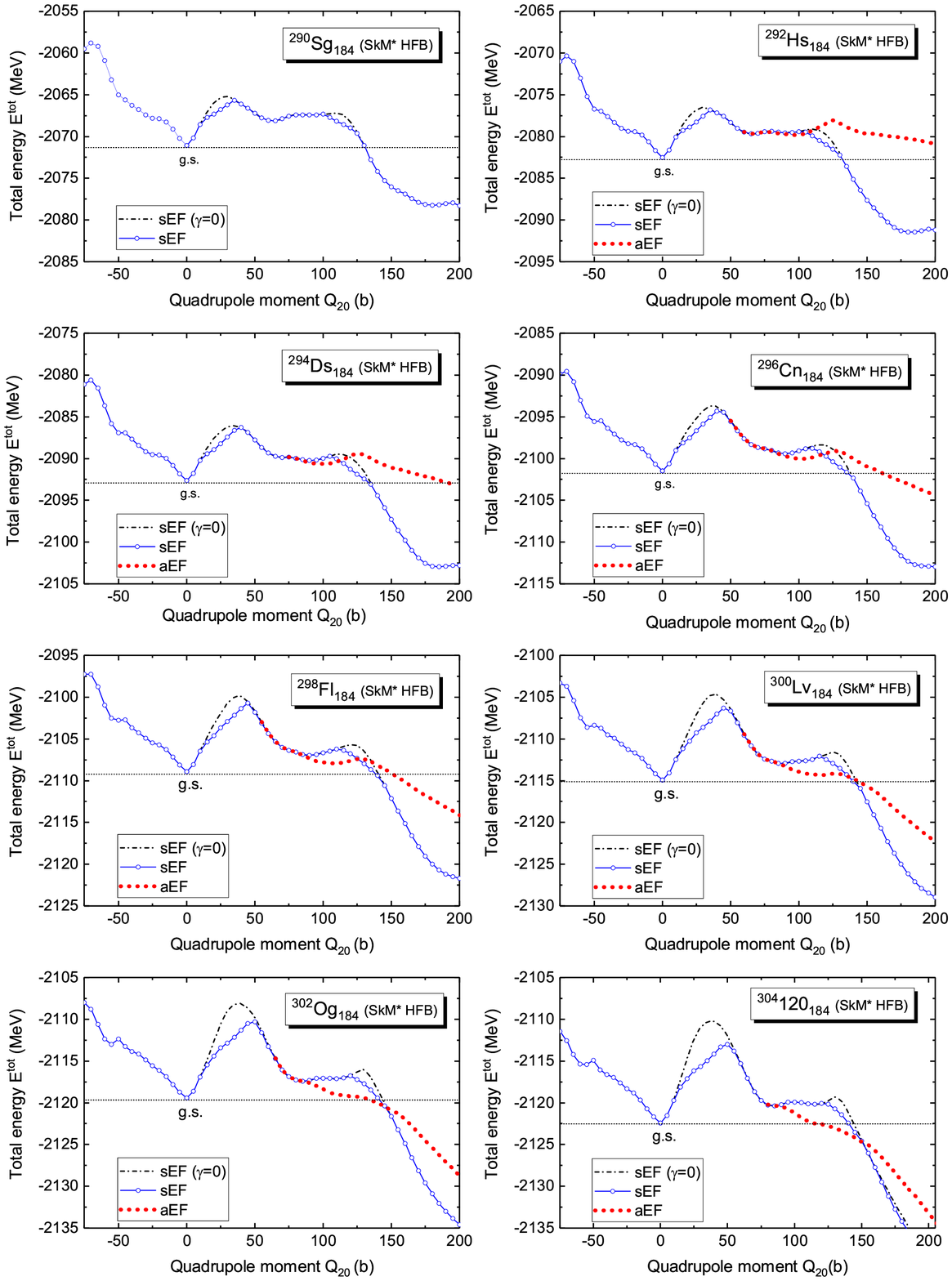}
\includegraphics[width=0.65\textwidth]{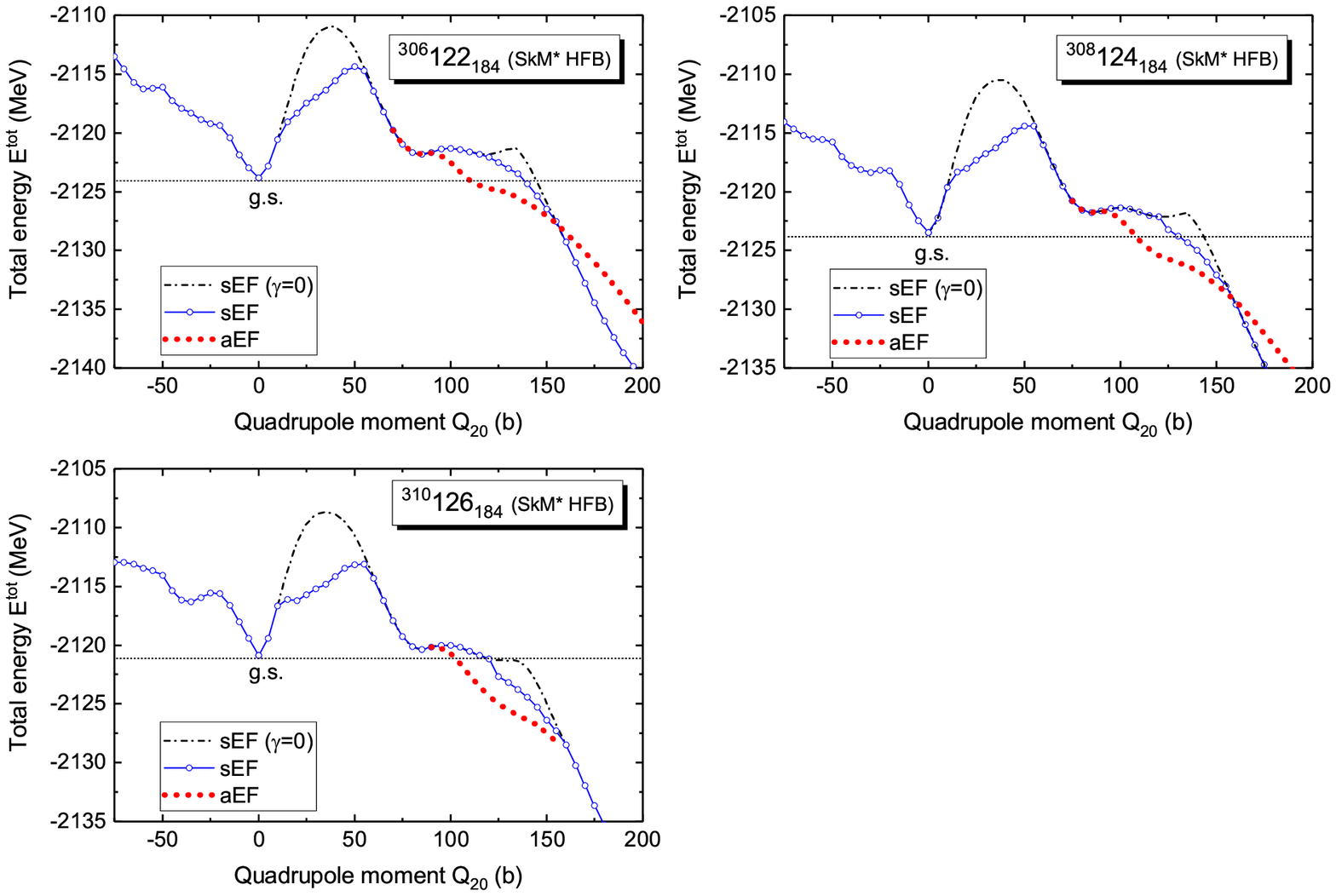}
\caption{\label{fig:3} (Color online) The same as Fig.~\ref{fig:1} and \ref{fig:2}, but for
even-even $N=184$ isotones with number of protons from $Z=106$ to $126$.
}
\end{center}
\end{figure}

\begin{figure}[htb]
\begin{center}
\includegraphics[width=0.75\textwidth]{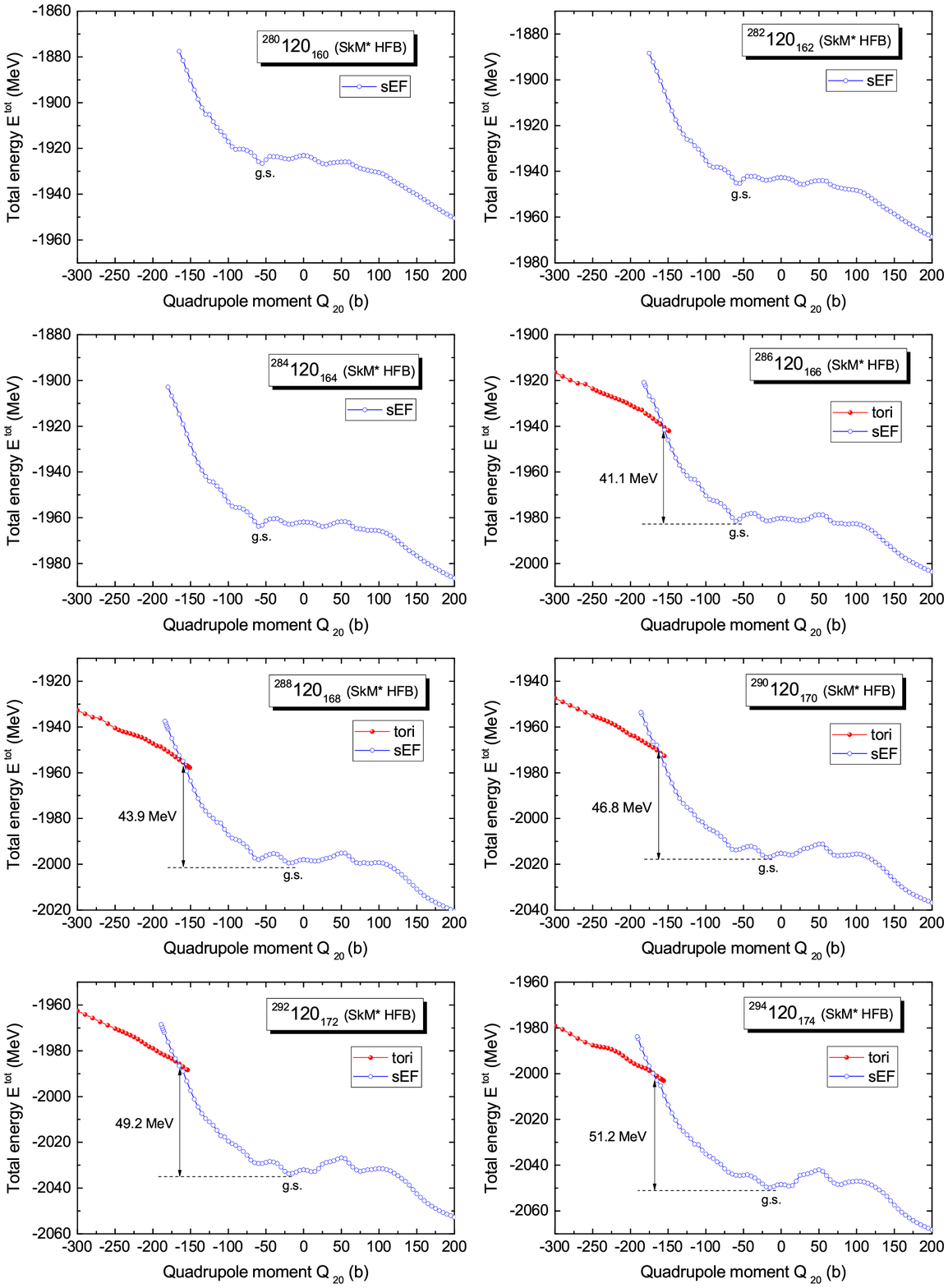}
\caption{\label{fig:5} (Color online) Total HFB energy of even-even $Z=120$ isotopes
with $N=160$ to $174$ as a function of the quadrupole moment.
The open circular points (blue color) show the symmetric elongated fission (sEF)
pathways. The nuclear matter density distributions with toroidal shapes appear at
the region of large oblate deformation $Q_{20} \lesssim - 160$ b (red solid circles).
}
\end{center}
%\vspace*{-0.6cm}
\end{figure}

The axially symmetric sEF ($\gamma =0^{\circ}$) fission paths are marked by the dash-dot
thin curves. The differences between the total HFB energies calculated along sEF and sEF with
$\gamma =0^{\circ}$ paths indicate dependence of fission barriers on triaxiality. Furthermore,
Figs.~\ref{fig:1} and \ref{fig:2} show that with increasing number of neutrons barrier
heights increase and reach a maximal value of $\sim 10$ MeV for $N=180$ and $182$.

\begin{figure}[htb]
\begin{center}
\includegraphics[width=0.75\textwidth]{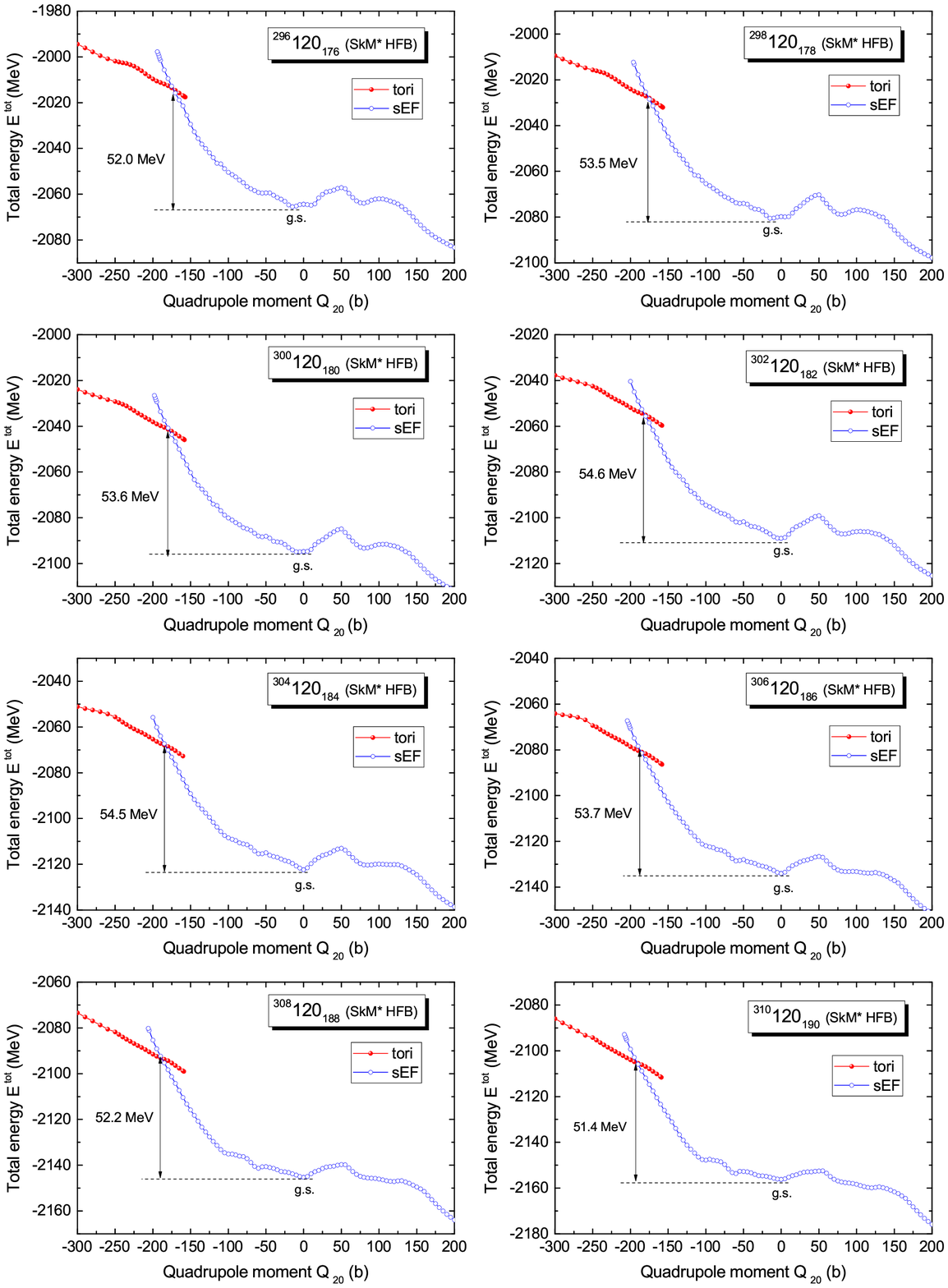}
\caption{\label{fig:6} (Color online) The same as Fig.~\ref{fig:5}, but for number of
neutrons from $N=176$ to $190$.
}
\end{center}
%\vspace*{-0.7cm}
\end{figure}

\begin{figure}[htb]
\begin{center}
\includegraphics[width=0.65\textwidth]{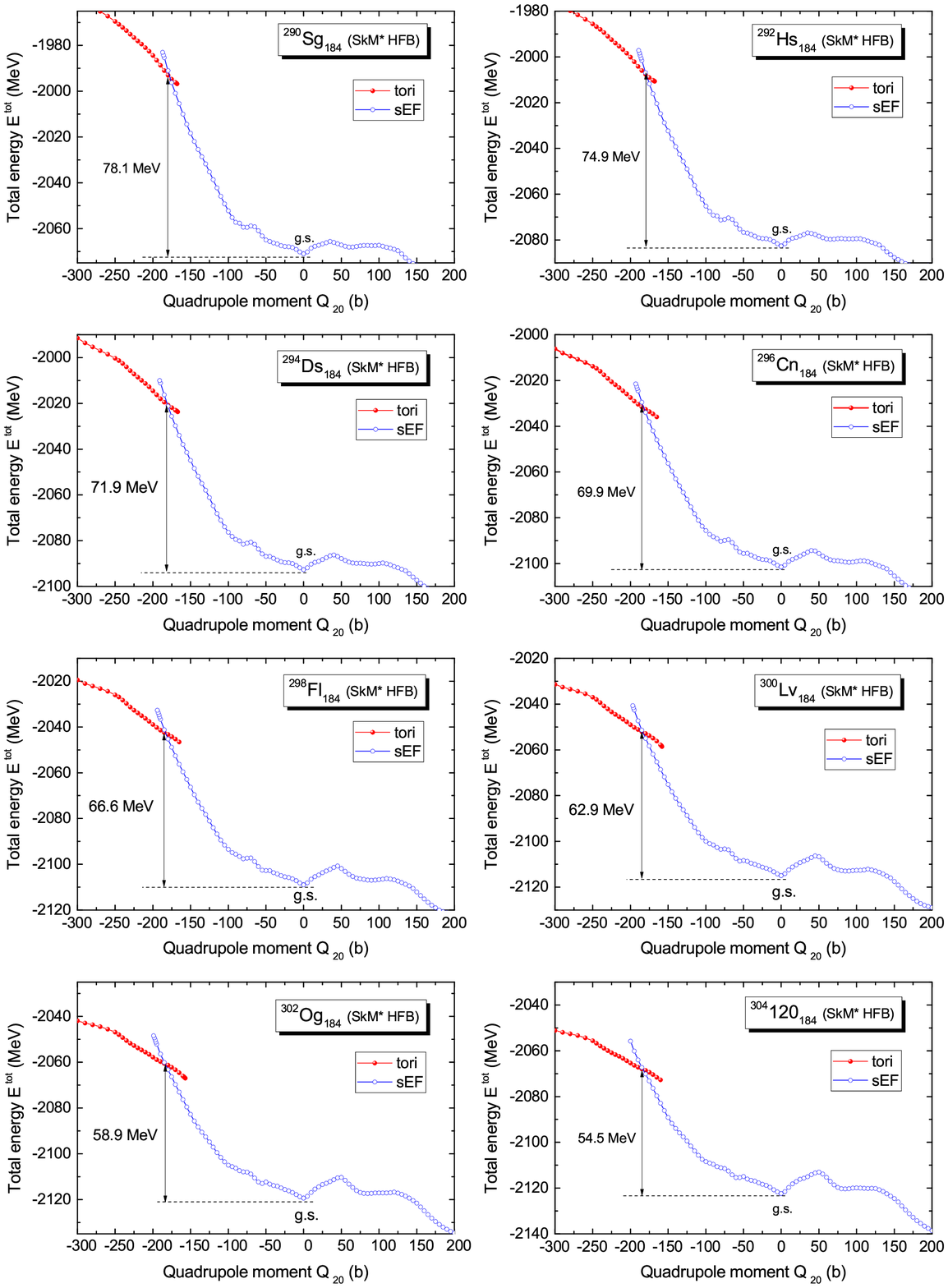}
\includegraphics[width=0.65\textwidth]{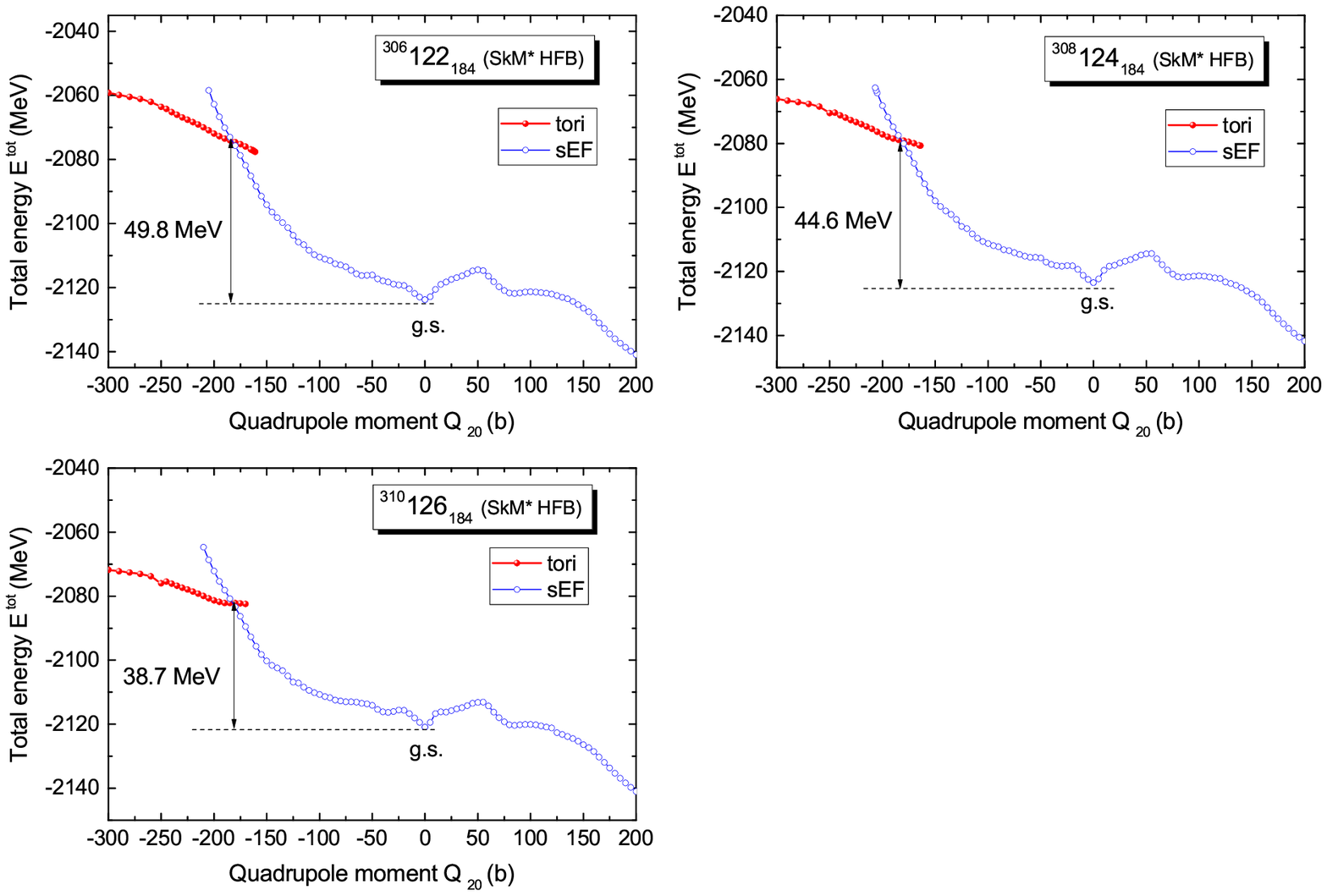}
\caption{\label{fig:7} (Color online) The same as Figs.~\ref{fig:5} and \ref{fig:6}, but for
even-even $N=184$ isotones with number of protons from $Z=106$ to $126$.
}
\end{center}
%\vspace*{-0.3cm}
\end{figure}

For neutron-deficient isotopes with number of neutrons $N$ from $160$ to $166$,
the ground state lies in the region of $Q_{20}\sim -50 $ b. This means that
these nuclei are super-oblate-deformed, see Ref.~\cite{Hee15}.
For the next group of six nuclei, i.e. with number of neutrons from $N=168$ to $178$,
there exist two minima, the ground state minimum in the region of $Q_{20}\sim - 25$ b and the
second one in the region of $Q_{20}\sim + 25$ b. It indicates the transitional nature of the nuclei.
For next nucleus, with $N=180$ neutrons, we have another situation, in which the barrier
between two minima disappears. This flat-bottom spherical potential allows the mixture
of oblate, spherical, and prolate shapes at the ground state, and can be viewed as the
E(5) critical-point solution \cite{Iach00} in the interacting boson approximation.
All of the next isotopes (with $N\geq 182$) have spherical shapes in the ground state.

Fig.~\ref{fig:3} presents the total HFB energy but for even-even superheavy $N=184$ isotones
with number of protons from $Z=106$ (Seaborgium) to $126$. All of them have spherical ground
state minima and the double-humped fission barriers (with inner and outer maxima). Moreover,
one can see that with the increase of protons number $Z$ the inner-barrier height rises and
for $Z=114$ to $124$ this barrier reaches the maximum value of $\sim 10$ MeV (taking into
account the effect of the lowering of the barrier height due to triaxiality).
Each of the examined isotones, except the Seaborgium, exhibit two fission paths: sEF and aEF.
For $Z\geq 116$ the aEF paths have substantially lower outer-barriers, which means that
these isotones favor the asymmetric fission path.

The total HFB energy surfaces in Figs.~\ref{fig:1} - \ref{fig:3} are presented for
the quadrupole moment $Q_{20}\geq -75$ b only. If the magnitude of oblate $Q_{20}$ deformation
increases the oblate spheroidal shapes of nuclei alter to the biconcave disc shapes and for
even greater oblate deformations a new family of toroidal shapes emerges, see Fig.~2 in
Ref.~\cite{Sta17}.
In Figs.~\ref{fig:5} and \ref{fig:6}, we explore the shape of the energy surface of even-even
$Z=120$ isotopes with $N=160$ to $190$ in the extremely oblate configuration by increasing
the magnitude of the constrained (negative) quadrupole moment ($Q_{20}\geq -300$ b).
For all isotopes with $N\geq 166$, the figures show how the energy surfaces vary as the shape
makes a transition from biconcave disc to the toroidal shape at the region of large oblate
deformation $Q_{20} \lesssim - 160$ b.

In addition, Figs.~\ref{fig:5} and \ref{fig:6} present the energy difference, $\triangle E$,
between the ground state and the energy point where two topological solutions (biconcave and
toroidal) have the same energy. The value of $\triangle E$ increases with the increasing
number of neutrons until the isotope $^{302}120_{182}$, with the value $\triangle E= 54.6$ MeV,
and then it starts to decrease slowly.

Fig.~\ref{fig:7} shows the same as Figs.~\ref{fig:5} and \ref{fig:6}, but for even-even $N=184$
isotones with number of protons from $Z=106$ to $126$. One can see, that when the number of
protons increases the energy difference, $\triangle E$, decreases rapidly, from 78.1 to 38.7 MeV,
and furthermore, the toroidal energy curves become oriented more horizontally.

\section{Toroidal High-Spin Isomers}

From the results in Figs.~\ref{fig:5} - \ref{fig:7} for the $Z$=120 isotopes and the
$N$=184 isotones under consideration, it is important to note that the
total energy curves in the toroidal configurations lie on a slope as a
function of $Q_{20}$, and there is no energy minimum there.
So without an angular momentum, these toroidal configurations are
unstable and have a tendency to return to a sphere-like
geometry. However, the stability of a toroidal nucleus may be enhanced
if it possesses an angular momentum about the symmetry axis
$I_{z}$. Because a quantal axially-symmetric toroid cannot
collectively rotate about its symmetry axis, only non-collective
rotations, such as those arising from particle-hole excitations are
possible in quantum mechanics.

\begin{figure}[htb]
\begin{minipage}[b]{0.47\linewidth}
\centering
\includegraphics[width=\linewidth]{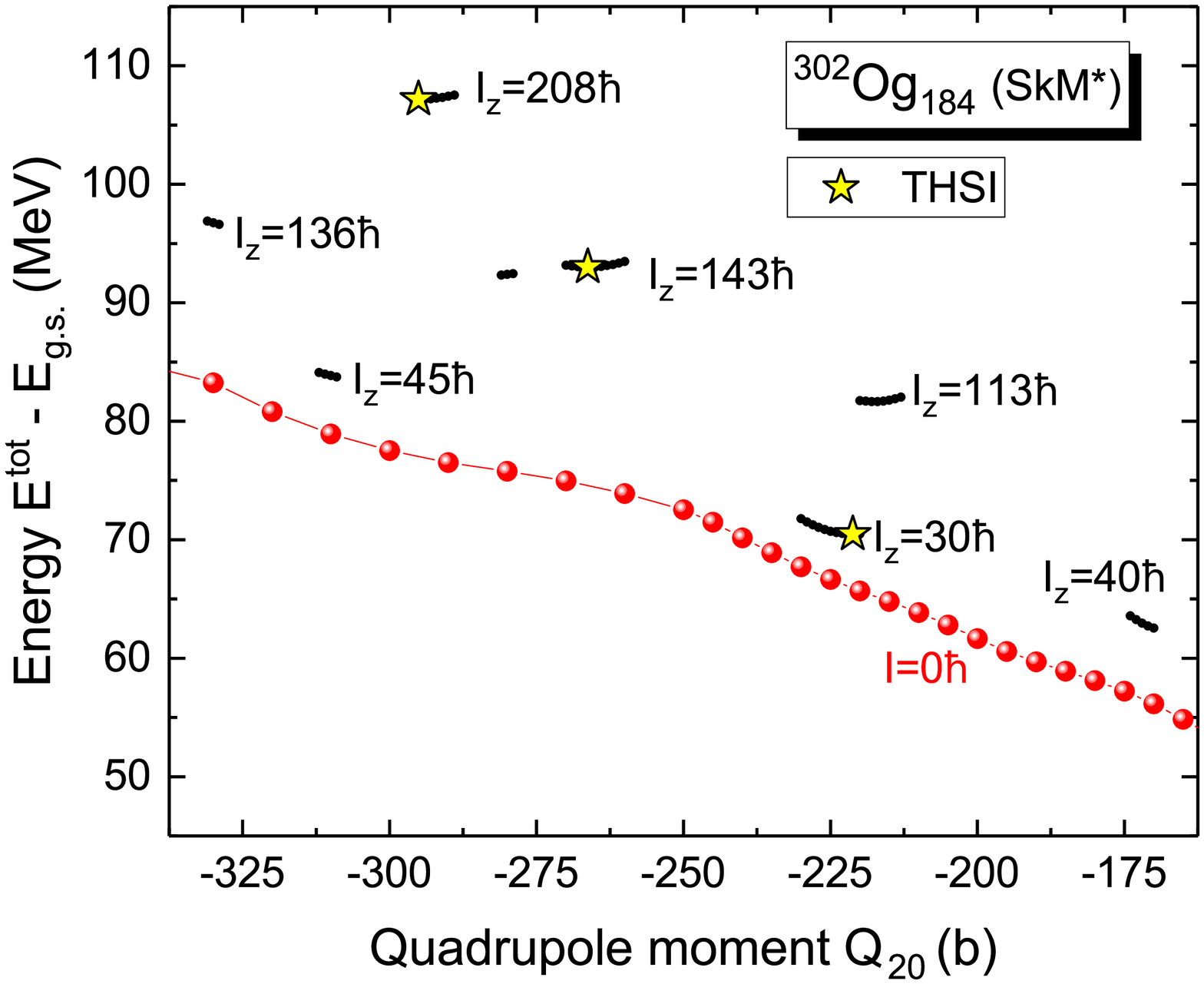}
\end{minipage}%
\hspace*{0.40cm}
\begin{minipage}[b]{0.47\linewidth}
\centering
\includegraphics[width=\linewidth]{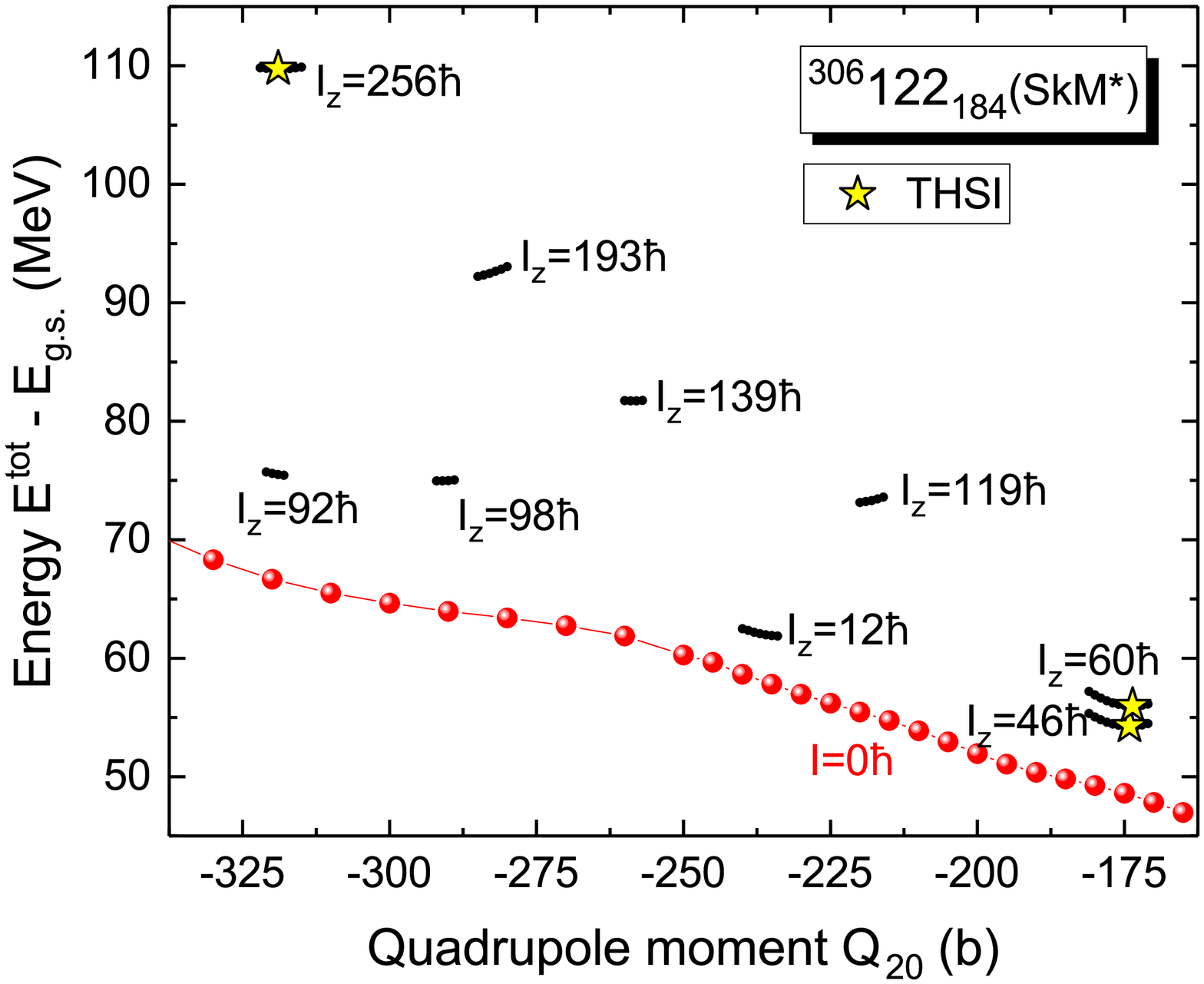}
\end{minipage}
\caption{\label{fig:9-10} (Color online) The deformation energies of $^{302}$Og$_{184}$ (left panel)
and $^{306}122_{184}$ (right panel) as a function of the quadrupole moment $Q_{20}$ for different
aligned angular momenta $I=I_{z}$. The locations of the toroidal high-spin-isomers (THSIs)
$^{302}$Og$_{184}(I_{z}=30, 143, 208\hbar)$ and $^{306}122_{184}(I_{z}=46, 60, 256\hbar)$
are indicated by star symbols.
All deformation energies are measured relative to the energy of the spherical ground state.
}
\end{figure}

\begin{figure}[htb]
\begin{minipage}[b]{0.47\linewidth}
\centering
\includegraphics[width=\linewidth]{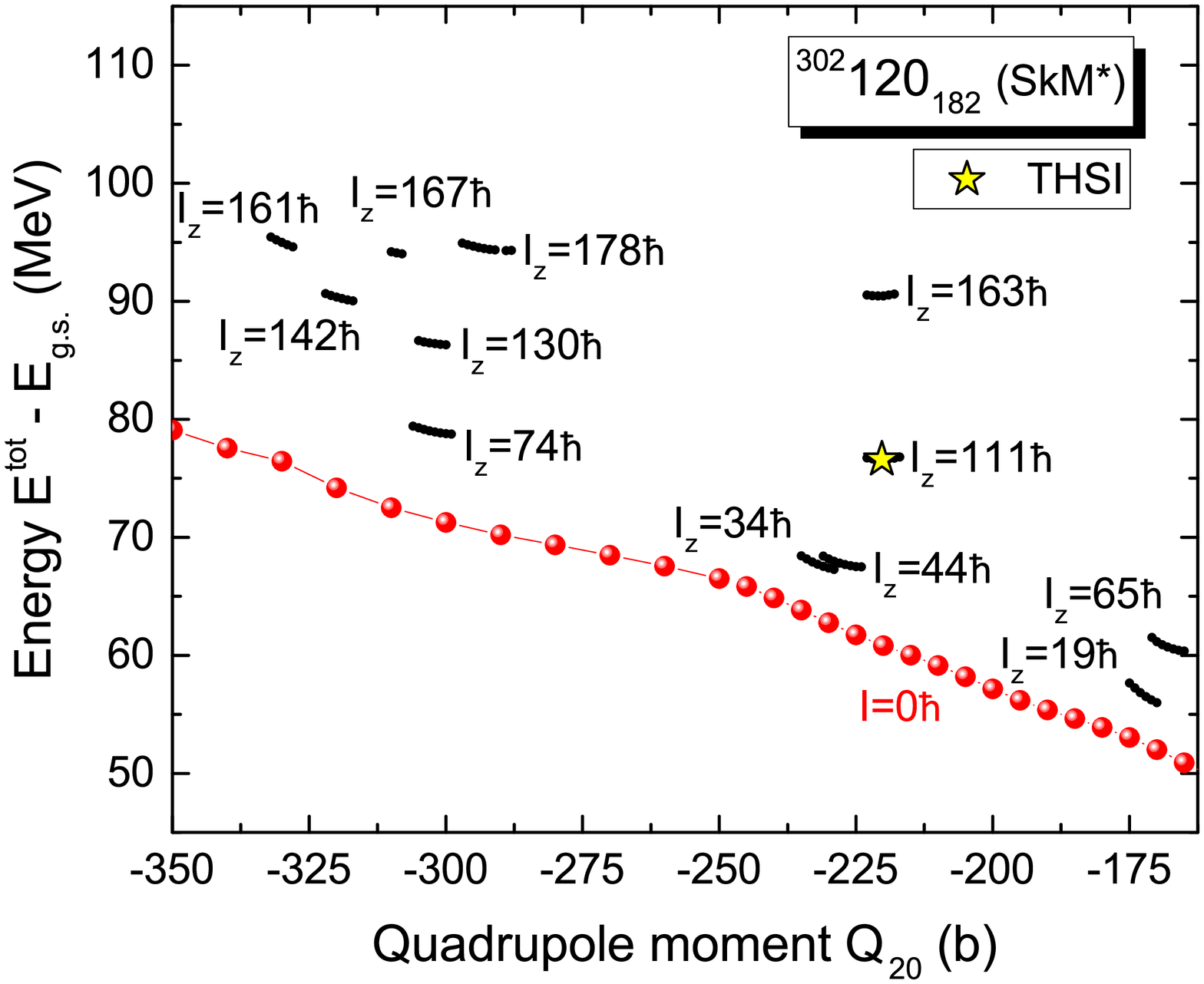}
\end{minipage}%
\hspace*{0.40cm}
\begin{minipage}[b]{0.47\linewidth}
\centering
\includegraphics[width=\linewidth]{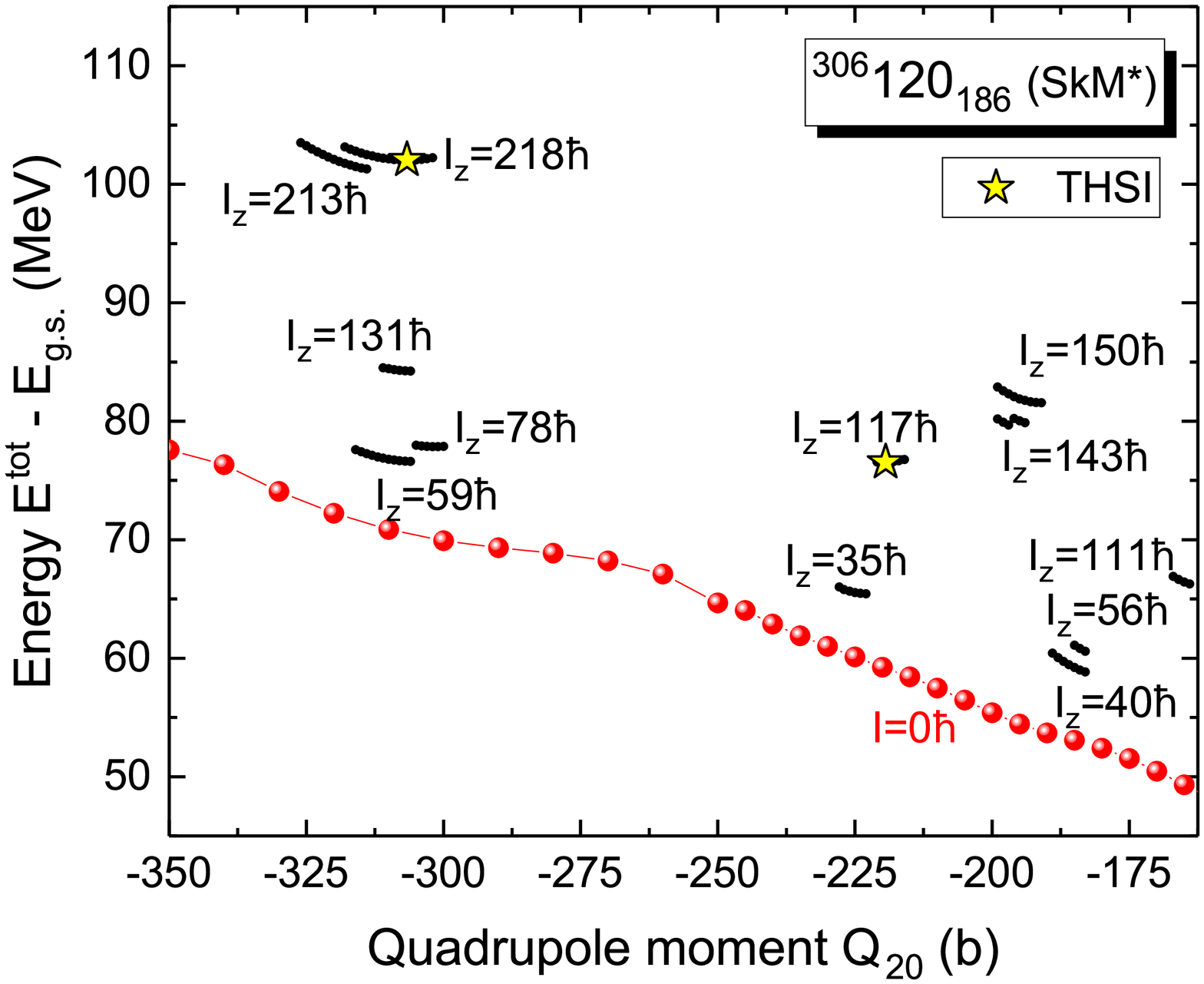}
\end{minipage}
\caption{\label{fig:11-12} (Color online) The same as Fig.~\ref{fig:9-10}, but for
$^{302}120_{182}(I_{z}=111\hbar)$ (left panel) and $^{306}120_{186}(I_{z}=117, 218\hbar)$
(right panel) THSIs.
}
\end{figure}

There are two equivalent ways to construct a high-spin state with spin aligned along
the symmetry axis, as presented previously in \cite{Sta17}, where we found THSI nucleus
$^{304}$120$_{184}$ with $I_{z} =81$ and $208\hbar$.
Using similar methods to explore neighboring even-even nuclei in
the $(Z,A)$ space, we obtain the energy curves of toroidal density
distributions with different angular momentum components $I_z$ aligned
along the symmetry axis. In Fig.~\ref{fig:9-10} we show sections of the
deformation energy surfaces of isotones $^{302}$Og$_{184}$ and
$^{306}122_{184}$ as a function of quadrupole moment $Q_{20}$ for
different aligned angular momenta.

We find that while many energy curves for different aligned angular
momenta do not possess an energy minimum, there are specific aligned
angular momenta for which the energy curves show localized energy
minima. These energy minima are indicated by the star symbols in
Fig.~\ref{fig:9-10} for THSI $^{302}$Og$_{184}$ with $I_{z}=30$,
143, 208$\hbar$, and $^{306}$122$_{184}$ with $I_{z}=46$, 60,
256$\hbar$.
The next Fig.~\ref{fig:11-12} represents the same situation as Fig.~\ref{fig:9-10},
but for isotopes $^{302}120_{182}$ and $^{306}$120$_{186}$ for which THSIs
exist for $I_{z}=111\hbar$ and $I_{z}=117$, 218$\hbar$, respectively.

The properties of found THSI states are tabulated in Table~\ref{tab:1},
where $\hbar\omega_{z}$ is the Lagrange multiplier (cranking frequency) and
$E^{*}$ denotes the energy measured relative to the energy of the spherical
ground state.

\begin{table}[htb]
\begin{center}
\caption{\label{tab:1} Properties of the toroidal high spins isomers (THSIs) of
$^{302}$Og$_{184}$, $^{306}$122$_{184}$,  $^{302}$120$_{182}$, and $^{306}$120$_{186}$. }
\vspace*{0.3cm}
\begin{tabular}{r l c c c c}
\hline
\hline
$Z$ & $N$ & $I_{z}$=$I_{z}^{proton}$+$I_{z}^{neutron} [\hbar]$ &Q$_{20} [b]$ & $\hbar\omega_{z} [MeV]$  & $E^{*} [MeV]$\\ [0.2ex]
\hline
118 & 184 & 30 = 15 + 15 & -221.2 & 0.05 & 70.5 \\
    &     & 143 = 59 + 84 & -266.3 & 0.20 & 93.0 \\
    &     & 208 = 79 + 129 & -295.1 & 0.30 & 107.2 \\
\hline
%\hline
122 & 184 & 46 = 14 + 32 & -174.1 & 0.10 & 54.3 \\
    &     & 60 = 14 + 46 & -173.6 & 0.15 & 56.0 \\
    &     & 256 = 98 + 158 & -318.9 & 0.32 & 109.7 \\
\hline
%\hline
120 & 182 & 111 = 44 + 67 & -220.3 & 0.20 & 76.6 \\
\hline
%\hline
120 & 186 & 117 = 40 + 77 & -219.4 & 0.20 & 76.5 \\
    &     & 218 = 79 + 139 & -306.7 & 0.30 & 102.0 \\
\hline
\hline
\end{tabular}
\end{center}
\end{table}

\section{Conclusions and Discussions}

We examine here properties of the superheavy nuclei of the even-even
$Z=120$ isotopes and $N=184$ isotones. The obtained results for $Z=120$
isotopes show the change of position of the ground state with the increase of
the number of neutrons from $N=160$ to $190$. The most neutron-deficient
isotopes form the group of the super-oblate-deformed nuclei, for $N=168$ to $178$
there exist the transitional region of nuclei with two oblate and prolate minima
separated by small barrier. This barrier disappears for $N=180$ and the
flat-bottom potential allows the mixture of oblate, spherical, and prolate
shapes at the ground state. Finally, all next isotopes with $N\geq 182$
are spherical in their ground states. In the case of $N=184$ isotones
with spherical ground states and the double-humped fission barriers,
it is found that with the increase of the number of protons the inner-barrier
height rises and reaches value of $\sim 10$ MeV for $Z=114$ to $124$.
Moreover, for majority of the $Z=120$ isotopes and $N=184$ isotones
one can observe competition between two paths leading to fission:
one with the symmetric elongated fragments (sEF) and second with
asymmetric elongated fragments (aEF).

Focusing our attention on the extremely oblate region, we search for
toroidal high-spin isomers in the neighborhood of $^{304}$120$_{184}$
where toroidal high spin isomers have been located previously in
theoretical calculations \cite{Sta17}. We find that the neighboring
even-even $N=184$ isotone with $Z=118$ and $122$, as well as the $Z=120$
isotopes with $N=182$ and $186$, also possess toroidal high spin
isomers at various angular momentum components and quadrupole moments.
The occurrence of toroidal high spin isomers may appear to be rather
common in the superheavy nuclei region.

\center {\bf Acknowledgement}

The research was supported in part by the Division of Nuclear Physics,
U.S. Department of Energy under Contract DE-AC05-00OR22725.

%uncomment the following lines to place a figure
%\begin{figure}[htb]
%\centerline{%
%\includegraphics[width=12.5cm]{Fig1}}
%\caption{Plot of ...}
%\label{Fig:F2H}
%\end{figure}

\end{document}